\documentclass[11pt,letterpaper,onecolumn,final]{article} % correct font and page size

\usepackage[]{cite} % references package
\usepackage[hmargin=0.75in,vmargin=0.75in]{geometry} % 0.75 inch margins
\usepackage{graphicx} % allows us to use more types of pictures
\usepackage{wrapfig} % allows us to wrap text around figures
\usepackage[format=hang,labelfont=bf,position=bottom,small]{caption} % caption formatting (bold labels, hanging indent, 10pt font)
\usepackage[nice]{nicefrac} % better fraction styling
\usepackage{amsmath} % better math typesetting
\usepackage{multirow} % allows better formatting in tables

\usepackage[T1]{fontenc}
\usepackage[scaled]{helvet}

\usepackage{titlesec}
\titleformat{\section}{\large\bf\scshape\raggedright}{}{0em}{}

\begin{document}

\begin{center}

% Title
\Large
\textbf{\textsc{
Ab Initio Discovery of Novel Crystal Structure Stability in Barium and Sodium-Calcium Compounds under Pressure
}}

% Authors (primary is underlined)
% followed by affiliations
\normalsize 
\vspace{2ex}
\underline{Joshua A. Taillon}, William W. Tipton, and Richard G. Hennig\\
Department of Materials Science and Engineering, Bard Hall, Cornell University, Ithaca, NY 14853
\vspace{2ex}

% Abstract minipage
\begin{minipage}[c]{5.5in}
\raggedright
Group I/II materials exhibit unexpected structural phase transitions at high pressures, providing potential insight into the origins of elemental superconductivity.  We present here a computational study of elemental barium and binary sodium-calcium alloys to identify both known and unknown phases of barium under pressure, as well as stable high-pressure compounds in the immiscible Na-Ca system.  To predict stability, we performed density functional theory calculations on randomly generated structures and evolved them using a genetic algorithm.  For barium, we observed all of the expected phases and a number of new metastable structures, excluding the incommensurate Ba-IV structure.  We also observed a heretofore unreported structure ($\alpha$-Sm) predicted to be the ground state from 28-46 GPa.  In the Na-Ca system, we demonstrate feasibility of our search method, but have been unable to predict any stable compounds. These results have improved the efficacy of the genetic algorithm, and should provide many promising directions for future work.
\end{minipage}

\end{center}

% main text
\raggedright
\section{Introduction:}

Without question, one of the most pressing challenges facing our society is how to more efficiently utilize energy.  One exciting solution is the development of superconducting materials,  but unfortunately, no materials have been found to date that exhibit superconducting behavior above 133 K, and the 20-year search for ``room-temperature'' superconductors has proved fruitless.  The highest-temperature superconducting materials have very complex cuprate structures, making these materials expensive to produce and difficult to characterize~\cite{mourachkine2004}, and there is still no agreed-upon microscopic model that can describe high-temperature superconductivity~\cite{leggett2006}.  As such, there is still a need for fundamental studies investigating superconductive behavior and under what conditions it develops.

Simple non-transition metals can display remarkable electronic and structural properties at high pressures and have become a popular superconductivity research alternative to cuprates. These Group I and II elements are ``simple'' due to their lack of \emph{d}-orbital outer shell electrons. While ordinary under ambient conditions, when placed under pressure they can adopt complex structures, exhibiting superconductivity and unexpected electronic transitions~\cite{degtyareva2010}.  Instead of attempting to find the highest T$_\textsf{c}$ possible, many high-pressure studies on these elements have aimed to investigate what features of a material give rise to this behavior.  A few studies have also investigated the effects of combining simple elements into compounds, finding that superconductivity can develop under high pressures in these systems as well~\cite{feng2008,xie2010}.

Experimental studies at high pressures are difficult, and as a result computational methods such as density functional theory (DFT) are widely used as an efficient and accurate way to predict interesting results and provide a sound theoretical basis for experimental studies.  DFT calculates the internal energy of an atomic system from its charge density using periodic boundary conditions and pseudopotentials to model interactions between atoms at zero temperature.  It allows \textit{ab initio} determination of structure stability based upon the pressure dependent enthalpy. For a detailed explanation of the theory, see~\cite{martin2004}.  Nearly all the simple metals have been shown to display some sort of interesting behavior at high pressures, but three in particular, barium, calcium, and sodium, serve as the basis of our study. 

\begin{table}[t]
\centering
\includegraphics[width=.7\textwidth]{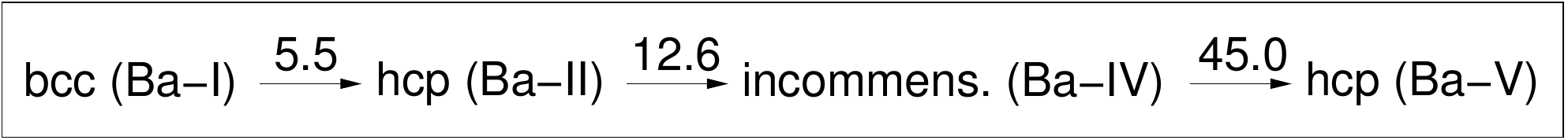}
	\caption[Phase Transformation Sequence of Barium]{Experimental pressure-induced phase transitions for barium at room temperature, given in GPa~\cite{nelmes1999}.}
	\label{tab:BaPhaseTrans}
\end{table}

A number of experimental and computational studies have focused on barium, attempting to fully characterize its phase space up to 90 GPa.  The known pressure-induced phase transformations (at room temperature) are displayed in Table~\ref{tab:BaPhaseTrans}.  Two features of barium's high pressure response are worth further investigation.  First, the hexagonal close packed phase is reentrant at high pressures. Second, the moderate pressure Ba-IV phase is complex and incommensurate.  This means that while consisting of only one element, the structure contains two distinct overlapping unit cells, creating a host-guest structure that does not exhibit long range order in one direction.  In Ba-IV, the host unit cell is body-centered tetragonal while the guest cells are face-centered tetragonal.   These unit cells have the same repeat unit in the lateral plane, but their \textit{c} lattice parameters form an irrational ratio, making them impossible to model completely accurately using periodic boundary conditions and difficult to characterize experimentally.  The structure was only solved experimentally in the past decade and has seen only limited theoretical investigation, making it a prime subject for our study.   For more details about this structure, see~\cite{nelmes1999,reed2000,heine2000}.

Calcium and sodium also exhibit interesting behavior under high pressures, including transitions to insulating, superconductive, and complex incommensurate phases~\cite{neaton2001,oganov2010}.  These elements are fully immiscible at ambient conditions~\cite{CaNaPhase}, but no studies of the binary system at high pressures exist.  Previous work by Feng \textit{et al.}~found that the similar Li-Be alloy (also immiscible at ambient conditions) develops compounds at high pressures, which display extraordinary electronic properties~\cite{feng2008}.  Thus, a focus of this work is to determine if a similar phenomenon occurs in the Na-Ca system. 

\section{Experimental Methods:}

We introduce in this study a genetic algorithm which can predict not only elemental ground state structures, but also search the composition space of an alloy and predict stable compounds in a variety of environments~\cite{tipton2007}. The algorithm is based on evolutionary theory, which places genetic value on the features of a structure that minimize its enthalpy.  In our method, we randomly generate structures to create an initial parent population, relax them, and calculate their enthalpy.  The structures are then assigned an evolutionary value based on their relative enthalpies.  The lowest energy (highest value) structures are automatically promoted to the next generation, while the remaining structures for the generation are created using a combination of evolutionary techniques such as random mutation of atom positions and lattice vectors or the crossover of two parent structures to create an offspring. The algorithm progresses in this manner until convergence is reached and no further improvement is seen from one generation to the next, at which point the most stable structure has been found.  The algorithm is very efficient and is able to effectively sample the entire solution space without prematurely converging to a local minimum.  For more specific details about our method, see~\cite{tipton2007}. 

While the algorithm is compatible with variety of energy computation packages, we implemented the commercial Vienna Ab-initio Simulation Package (VASP),~\cite{vasp1,vasp2,vasp3,vasp4} using the generalized-gradient approximation with the Perdew-Berke-Ernzerhof (PBE) exchange correlation functional at zero temperature to model electron interactions~\cite{PBE,PBEerratum,PBExc}.  At the high pressures examined in this study, we expect core electrons to contribute to bonding.  Thus, we implemented an effectively all-electron calculation strategy to include core electron effects, using a plane wave basis set within the projector augmented wave method~\cite{PP2PAW,PAW}.  Structures were relaxed within VASP using the conjugate gradient algorithm to minimize energies.  We sampled the reciprocal space of our structures with a gamma-centered Monkhorst-Pack k-point grid  automatically generated by VASP.

\section{Results and Discussion:}

The first step of any computational study is to ensure that the calculations being performed are accurate and precise enough to be accurately repeated.  While default parameters were used for the structure search, our phase stability calculations required extremely precise results.   We converged the two relevant VASP input parameters, the plane wave energy cutoff (ENCUT), and the k-point mesh sampling density (KPOINTS length). In VASP, the density of k-points is given as a linear density, and defines the number of k-points per reciprocal lattice vector.  This parameter tends to be around 10 for wide-gap insulators and up to about 100 for \textit{d}-band metals.  Because the Fermi surface of \textit{d}-band metals tends to be much more complex than that of insulators and \textit{s}-band metals, many more k-points are required for an accurate Brillouin Zone integration.   

Both of these parameters were increased systematically to very high values, and the enthalpy of a static lattice calculation recorded at each value. Each parameter was increased until the difference between the enthalpy per atom at each point and the enthalpy per atom at the highest value of the parameter was less than 1 meV.  Because the values of these parameters depend on the pseudopotentials used and electronic properties of the elements, this process was repeated for each element in this study at four pressures ranging from 0 to 300 GPa. The criterion used ensures that our calculations are precise to within 1 meV and the results of our convergence calculations are reported in Table~\ref{tab:converge}. \vspace{6pt}

\begin{table}[h!]
\centering
\small
\begin{tabular}{ | c | c | c | }
\hline  \textbf{Element} & \textbf{ENCUT} [eV] & \textbf{KPOINTS length} [\AA$^{\mathrm{\textsf{-1}}}$] \\ 
\hline 	Barium & 400 & 45 \\ 
         Calcium & 500 & 40 \\ 
	 Sodium & 1175 & 40 \\ 
\hline 
\end{tabular} 
\caption{Summary of elemental convergence parameters.  When used in an alloy, the highest convergence values of each of the constituent elements must be used.  Thus, we used the sodium values in our Na-Ca alloy phase stability calculations.}
\label{tab:converge}
\end{table}

We performed our barium genetic algorithm search at seven pressures throughout the 300 GPa range investigated in this study.  The correct volume per atom was determined by relaxing an experimental unit cell at each pressure. Using this volume, a number of randomly generated structures were created at each pressure and used as the initial population for the genetic algorithm.  Performing this search at seven different pressures was adequate to discover all the experimentally found ground structures, excluding the incommensurate complex Ba-IV structure.  On average, the algorithm was allowed to run until six generations of approximately 24 structures each had been relaxed, and the best three structures were automatically promoted in each generation. Calculations during the search were performed using default VASP convergence parameters, which speeds the search and is adequate to find competitive structures.
  
The results of the barium structure search are summarized in Table~\ref{tab:BaGAresults}.  Once the algorithm finished developing structures, the space group and atomic positions of each structure were determined using the ISOTROPY program~\cite{ISOTROPY}.  At each pressure, we identified the unique low enthalpy or high symmetry structures that were to be compared in phase stability calculations.  We identified our structures using the catalog maintained by the United States Naval Research Laboratory~\cite{navalStructures}. Common structures are represented in our results by their prototype, while the less common are identified by their space group and number of atoms per conventional unit cell.

An interesting result from our structure search is that beyond 100 GPa, the same three phases were identified in each search.  This indicates that the phase space of barium at very high pressures condenses, and only close packed structures remain competitive.  As seen in later calculations, the close packed \textit{P$\overline{\mathrm{\sf{\textit{3}}}}$m1}-5 structure, found at 60 GPa, is also competitive at high pressures.  The symmetry distortions seen in the low pressure searches are suppressed in these extreme environments and the atoms are forced into the highest possible packing efficiency, which is possible in only a few very distinct structures.

\begin{table}[ht]
\centering
\small
	\begin{tabular}{|c||cccc|} \hline
		\textbf{Pressure [Gpa]} 	& \multicolumn{4}{|c|}{\textbf{Structures}}\\ \hline
		\multirow{2}{*}{0}	& $\alpha$-Uranium & \textit{P2$_\mathrm{\textit{1}}$/m}-6 & BCC & \textit{C2/m}-5 \\
				& FCC & BCT & Simple Hex. &  \\ \hline
		10		& HCP & $\alpha$-Uranium & \textit{Imm2}-4 & \textit{C2/m}-5 \\ \hline
		30		& HCP & \textit{Pnma}-4 & FCC &  \\ \hline
		60 		& $\alpha$-Samarium & HCP & $\alpha$-Uranium & \textit{P$\overline{\mathrm{\sf{\textit{3}}}}$m1}-5\\ \hline
		100~~200~~300	& HCP & $\alpha$-Samarium & FCC &  \\ \hline
	\end{tabular}
	\caption[Summary of Barium Genetic Algorithm Results]{Summary of barium genetic algorithm structure search results.  The structures listed above are identified by the prototypical structure, or by space group name and the number of atoms per conventional unit cell, and are listed in order of increasing enthalpy per atom.  The second row of 0 GPa structures were not the lowest in energy, but displayed enough high symmetry elements to warrant further investigation.  Of particular interest are the identical results for all three high-pressure searches.}
	\label{tab:BaGAresults}
\end{table}
  
\begin{wrapfigure}{r}{0.5\textwidth}
 \vspace{-24pt}
  \begin{center}
    \includegraphics[width=0.5\textwidth]{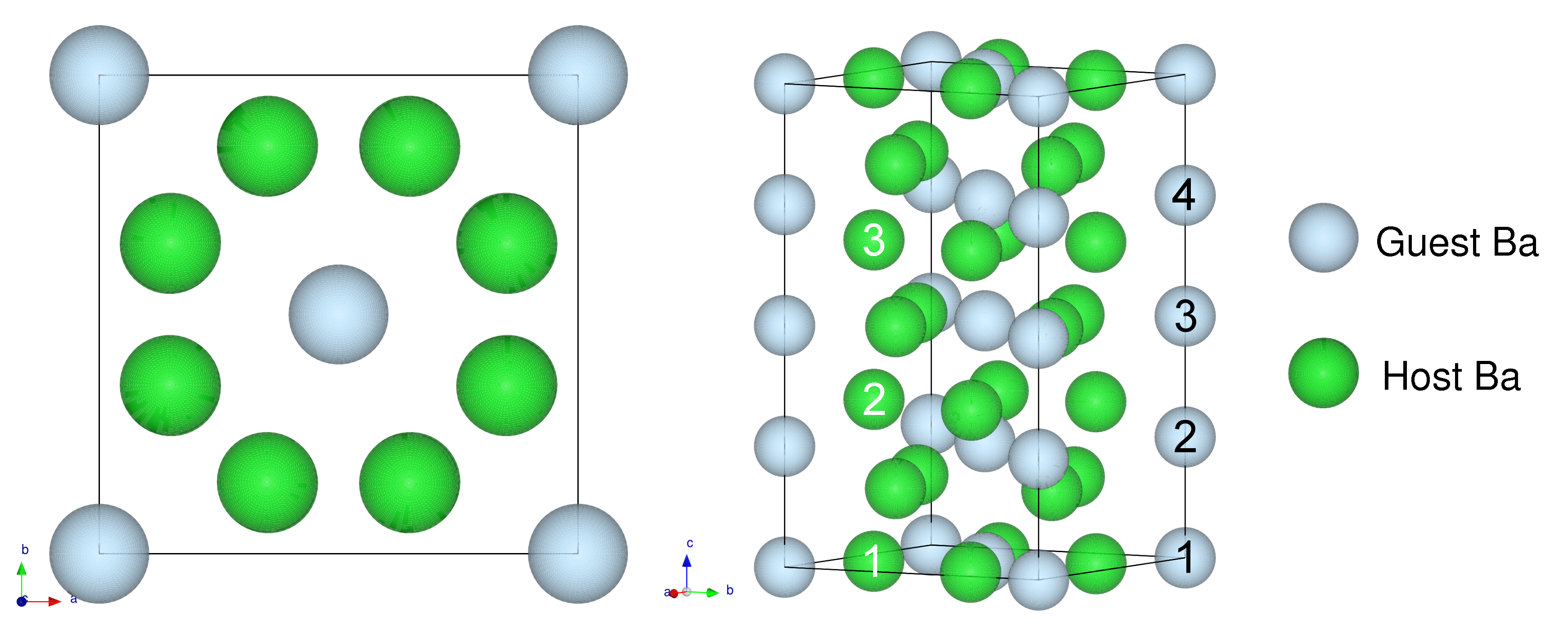}
  \end{center}
  \vspace{-10pt}
  \caption{4/3 Commensurate analog of Ba-IV.  This structure is in space group \textit{I4/mcm} with 32 atoms in Wyckoff positions \textit{4a}, \textit{4c}, \textit{8h} (x~=~0.35271), and \textit{16l}\\ (x~=~-0.35011 and z~=~-0.33503). Structure parameters are for the structure at 12.6 GPa. The three host cells and four guest cells are indicated by number (\textit{color online}).}
  \label{fig:BaIV}
  \vspace{-11pt}
\end{wrapfigure}

The complex Ba-IV phase presented a significant challenge to our algorithm.  Because it is incommensurate, this structure cannot be perfectly represented by periodic boundary conditions and must be approximated by a commensurate analog.  An appropriate analog will have a \textit{c$_\mathrm{\textsf{host}}$}/\textit{c$_\mathrm{\textsf{guest}}$} ratio similar to that of the experimental structure, which was found to be 1.388 at 12.6 GPa~\cite{nelmes1999}.  In their computational study of Ba-IV, Reed and Ackland used a 3/2 commensurate analog containing 11 atoms in 2 host cells and 3 guest cells (1.5 ratio).  While this was a reasonable approximation of the structure, it required a large (65 meV) correction to match experimental data~\cite{reed2000}.  Thus, this structure is unlikely to be discovered in our search.  The next reasonable analog is a 4/3 structure containing 32 atoms, used by Arapan \textit{et al.} in a computational study of the same structure in calcium and pictured in Fig.~\ref{fig:BaIV}~\cite{arapan2008}.  They predicted stability of this structure in calcium using similar methods, indicating that this is an appropriate analog to use in our study of barium.  Our search was limited to structures containing fewer than 15 atoms per unit cell in order to keep computation times reasonable, thus preventing this analog from being  discovered.  Our search was successful for all other phases of barium however, so we believe that a repeat search constrained to larger unit cells would be able to successfully identify the Ba-IV commensurate analog.

Once we characterized the genetic algorithm structures, we performed phase stability calculations  to determine which structures were stable as a function of pressure.  We relaxed each structure twice with the converged VASP parameters, using the relaxed geometry of the first calculation as the starting structure for the second, ensuring a complete relaxation.  We increased pressure gradually from 0 to 300 GPa, well beyond the 90 GPa limit seen in previous literature. In addition to structures found by the genetic algorithm, we included the 4/3 commensurate analog of the Ba-IV phase and the ideal-$\omega$ structure, which is seen in nearby elements such as Hf, Zr, and Ti.

In their paper, Reed and Ackland used~a~3/2 commensurate analog of Ba-IV, but included a large correction based on the strain energy required to force the incommensurate structure into this conformation.  Two particular aspects of their correction are worrisome.  First, they assume that the strain required to force the ratio from 1.388 to 1.5 (8\%) is completely elastic, which is unlikely.  Second, they assume a constant value of 150 GPa for the C$_{\textsf{33}}$ elastic constant.  We determined this constant (using a finite strain calculation within VASP) at a range of pressures from 10-15 GPa and found that it varied from 40-60 GPa, indicating a major flaw in Reed and Ackland's correction.  The commensurate analog we chose to use (pictured in Fig.~\ref{fig:BaIV}) is 4/3, meaning that it consists of four face centered tetragonal guest cells and three body centered tetragonal host cells.  The structure contains 32 atoms per unit cell and \textit{c}-axis ratio deviates only 4\% from the experimental. We expect this representation to be a much better approximation of the Ba-IV phase than the structure used by Reed and Ackland, eliminating the need for an energy correction.

\begin{figure}[b!]
\centering
\vspace{-10pt}
\includegraphics[width=\textwidth]{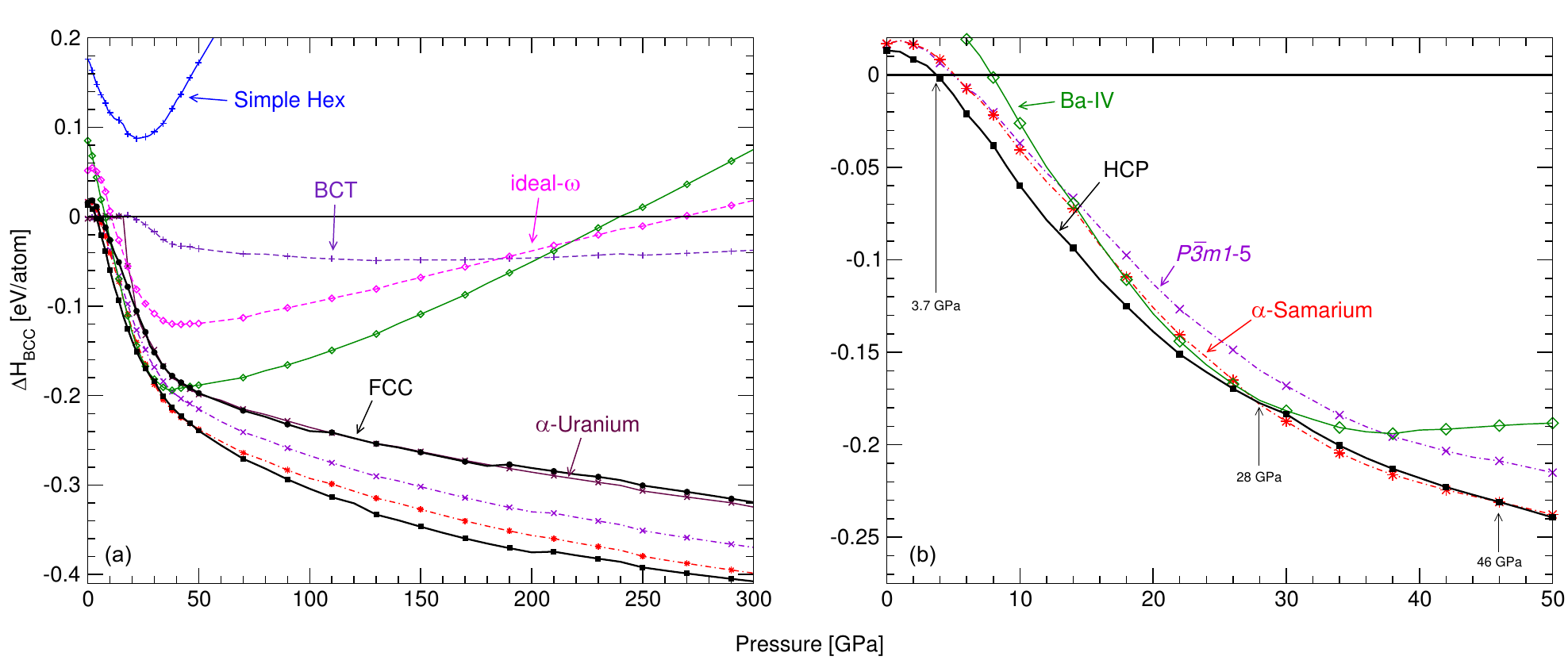}
\vspace{-10pt}
	\caption[Phase Stability Calculations]{Phase stability calculations for elemental barium as a function of pressure (a) over the entire range investigated, and (b) enlarged to the  0-50 GPa range, showing only competitive phases.  The data is plotted relative to the enthalpy of the BCC phase, and the structure that minimizes enthalpy at a given pressure is predicted to be stable at that pressure. Note: The \textit{P2$_\mathrm{\textit{1}}$/m}-6, \textit{C2/m}-5, \textit{Imm2}-4, and \textit{Pnma}-4 structures are omitted from this figure for clarity.  The first two were found to have negligible energy fluctuations around BCC and eventually relaxed into the FCC structure at high pressures.  The second two were found to be degenerate in energy with HCP over the entire pressure range (\textit{color online)}.}
	\label{fig:BaHvP}
\end{figure} 

The results of our phase stability calculations are shown in Fig.~\ref{fig:BaHvP}.  Notably, the HCP phase remains stable at all pressures above 46 GPa, and every structure other than simple hexagonal was found to be competitive at some pressure.  We plotted the data as enthalpy difference from the BCC phase rather than absolute enthalpy, causing the zero point energy contributions of each phase to cancel.  This means any enthalpy difference greater than our convergence criterion of 1 meV can be regarded as significant. A few results from our calculations are particularly interesting:

First, our calculations on the Ba-IV structure did not predict this phase to be stable at any pressure.  Our analog structure becomes degenerate with HCP around 28 GPa, but this is the closest approach to stability of the phase.  The energy values shown in Fig.~\ref{fig:BaHvP} include no correction term like that used by Reed and Ackland, but we observe that our analog structure is much more competitive than the one used in their study, requiring a correction of only 30 meV to match the experimental data, less than half that used previously. Consequently, using their correction formula of $ \Delta\textsf{E}=\nicefrac{\textsf{V}}{\textsf{4}}~\mathrm{\textsf{C}}_{\textsf{33}} \left[\textsf{1}-\nicefrac{\textsf{2}}{\textsf{3}}\times\nicefrac{\textsf{c}_\textsf{host}}{\textsf{c}_\textsf{guest}} \right]^\textsf{2} $, along with the structure parameters of our analog at 15 GPa ($\textsf{V}\textsf{ = }\textsf{31 \AA}^\textsf{3} \textsf{per atom}$, \textsf{C}$_{\textsf{33}}$ = \textsf{50 GPa}, and \nicefrac{\textsf{c}$_\textsf{host}$}{\textsf{c}$_\textsf{guest}$} = \nicefrac{4}{3}),  we find an energy correction of 29.86 meV per atom.  This correction gives transition pressures of 12 and 40 GPa, in very good agreement with experiment.  Although these values are accurate, we do not place much faith in this correction due to the assumptions it requires, namely, the unlikely pure linear elastic behavior and constant elastic properties.  

While our analog is much more competitive than that used by Reed and Ackland, it is still not stable, indicating that this structure experiences a destabilization at low temperature, and is likely not the true ground state at zero temperature.  It is thought that the stability of Ba-IV arises from the interaction of two distinct types of Ba atoms with different electronic character.  The host atoms display more \textit{s} character, while the guest atoms display more highly localized \textit{d} character~\cite{reed2000}. Thus, this structure can be viewed as an intermediate between the fully \textit{s}-like Ba-II phase and the fully \textit{d}-like Ba-V phase. It seems likely that some effect at low temperature causes a disruption in this transition, meaning the structure only becomes stable at room temperature.  It is possible however, that other modes of stabilization, not well-described by DFT, might lower the energy of this structure such that it becomes the ground state. Nevertheless, it is not uncommon for the lowest enthalpy ground state structures to disagree with room-temperature experimental phases~\cite{oganov2010}, and that is the conclusion indicated by our results.

\begin{wrapfigure}{l}{.4\textwidth}
 \vspace{-10pt}
    \includegraphics[width=0.4\textwidth]{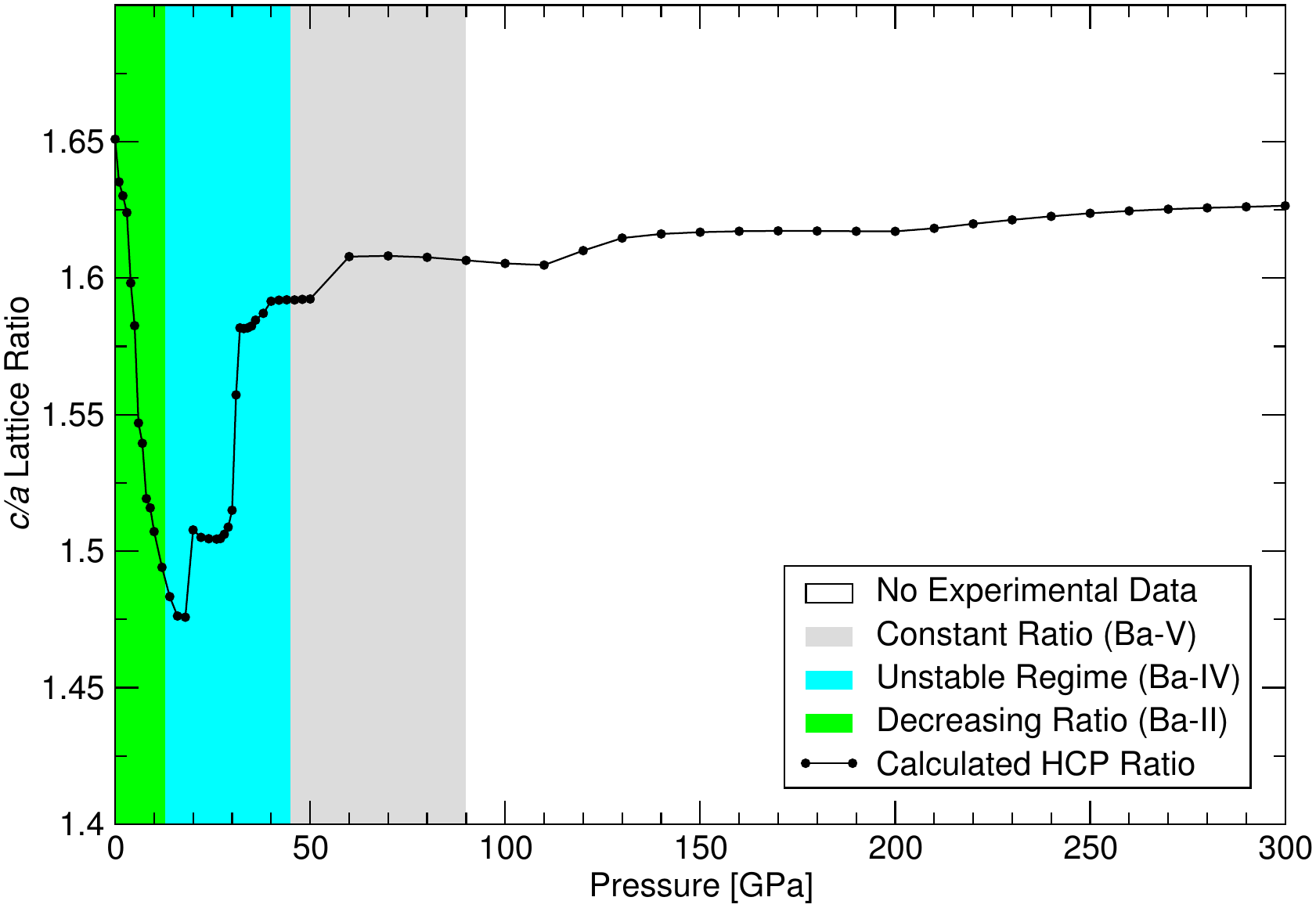}
  \vspace{-14pt}
  \caption{Calculated HCP \nicefrac{\textit{c }}{\textit{ a}} lattice ratios as a function of pressure. The shading represents the trends observed experimentally~\cite{reed2000}.}
  \label{fig:BaHCP}
  \vspace{-10pt}
\end{wrapfigure}The next interesting result from our calculation is the presence of a clear cusp in energy in the HCP phase at 30 GPa.  This points to a potential destabilization of the phase around this pressure, as well as a reentry of the phase at high pressures, as seen experimentally.  We performed additional analysis on the HCP phase and found that the \nicefrac{\textit{c }}{\textit{ a}} lattice ratio exhibits the same trends under pressure theoretically as seen experimentally (see Fig.~\ref{fig:BaHCP}).  As the pressure approaches 20 GPa, the lattice ratio falls drastically and the packing efficiency decreases, a distortion that includes an energetic penalty.  Throughout the experimentally unstable region from 13-45 GPa, the ratio increases, and remains relatively constant at higher pressures, approaching the ideal value of 1.633.  We found a deviation from experiment in our calculated BCC to HCP transition pressure.  Our value of 3.7 GPa is over 30\% below the experimental value of 5.5 GPa.  Qualitatively, we can justify this deviation because our calculations are at zero temperature.  As temperature is lowered, the BCC to HCP transition pressure is observed to decrease due to relative entropic contributions to the total energy~\cite{chen1988}.  This effect is not fully quantified however, so it is difficult to infer how low the transition shifts experimentally.

 \begin{wrapfigure}{r}{.5\textwidth}
\vspace{-30pt}
\centering
\includegraphics[width=.48\textwidth]{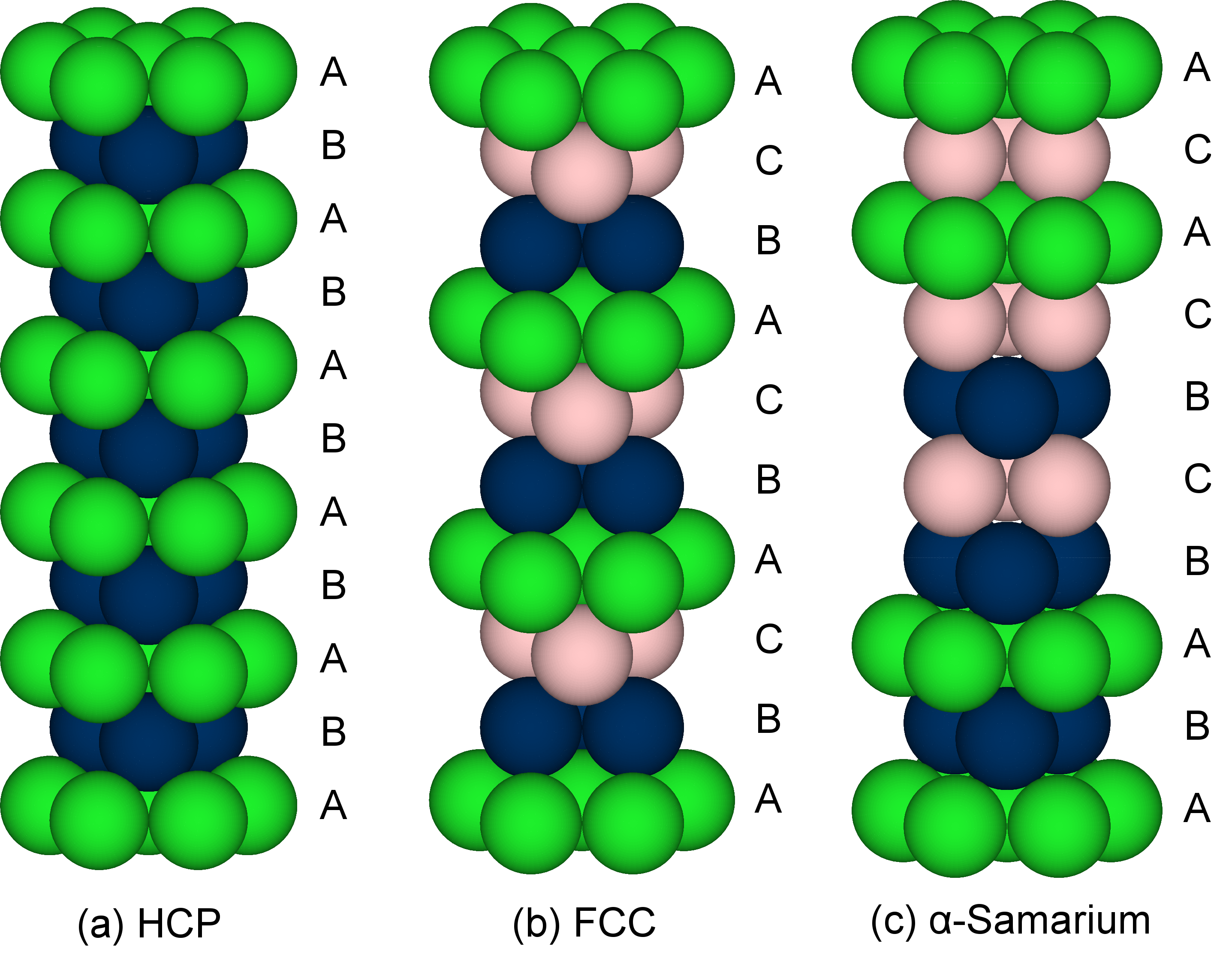}
\caption[Close Packed Structures of Barium]{Close packed structures of barium. (\textit{a}) HCP barium exhibits a simple ...AB... stacking pattern; (\textit{b}) FCC contains three layers of close packing in an ...ABC... pattern; (\textit{c}) $\alpha$-Sm displays a more complicated ...ABABCBCAC... nine-layer pattern.}
\label{fig:closepacked}
\vspace{-32pt}
\end{wrapfigure}
Finally, the most interesting result of our stability calculations involves the $\alpha$-Sm phase, which was discovered in our structure search.  This phase has not been reported in barium previously, but has been proposed as the ground state structure for lithium~\cite{overhauser1984}.  It is a close packed structure, similar to HCP and FCC, but has a nine layer stacking pattern, as seen in Fig.~\ref{fig:closepacked}. The formation of this phase is likely realized by the combination of energetically favorable stacking faults within the other stable close packed structures. The ends of the $\alpha$-Sm unit cell follow HCP stacking rules, but are offset by one interstice.  The transition between these two regions occurs in the center of the cell, introducing two head-to-head units of FCC-like stacking (...ABC-BCA...).  In this way, the $\alpha$-Sm phase appears to be a more stable intermediate between the two standard close packed phases.

The differences in stacking between these structures arise primarily due to electronic effects, and have been well-studied in both samarium and lithium.  A primary force behind the transition from FCC to HCP (and also $\alpha$-Sm) is the occupancy of the conduction band in each phase (the valence)~\cite{redfield1986}. This indicates that as the external pressure is increased in barium, there is a change in bonding and in the partial occupancies.  Specifically, the electrons in the \textit{s} band move to the \textit{d} band,  a transition observed in both the HCP and Ba-IV structures.  Thus, it is not unexpected to see this effect give rise to a new phase, and a further DFT analysis of the electronic properties, including the density of states, should prove fruitful in characterizing the transition between these structures, and the fundamental reasons for the $\alpha$-Sm phase stability.

We found the $\alpha$-Sm structure to be the ground state of barium from 28-46 GPa.  This determination is sensitive to the correct enthalpy values for Ba-IV, but we believe this structure to be the true ground state at 0K in this range.  The experimental prevalence of Ba-IV, as well as $\alpha$-Sm's similarity to both FCC and HCP means that this structure may never be observed experimentally. Its emergence in our calculations indicates that the same forces that lead to the destabilization of HCP (\textit{s} $\rightarrow$ \textit{d} electron transfer) might give rise to the $\alpha$-Sm structure.  Because of the similar high pressure response of other simple metals, this structure deserves an investigation in those elements as well.

\begin{wrapfigure}{r}{.5\textwidth}
\centering
\vspace{-12pt}
\includegraphics[width=.5\textwidth]{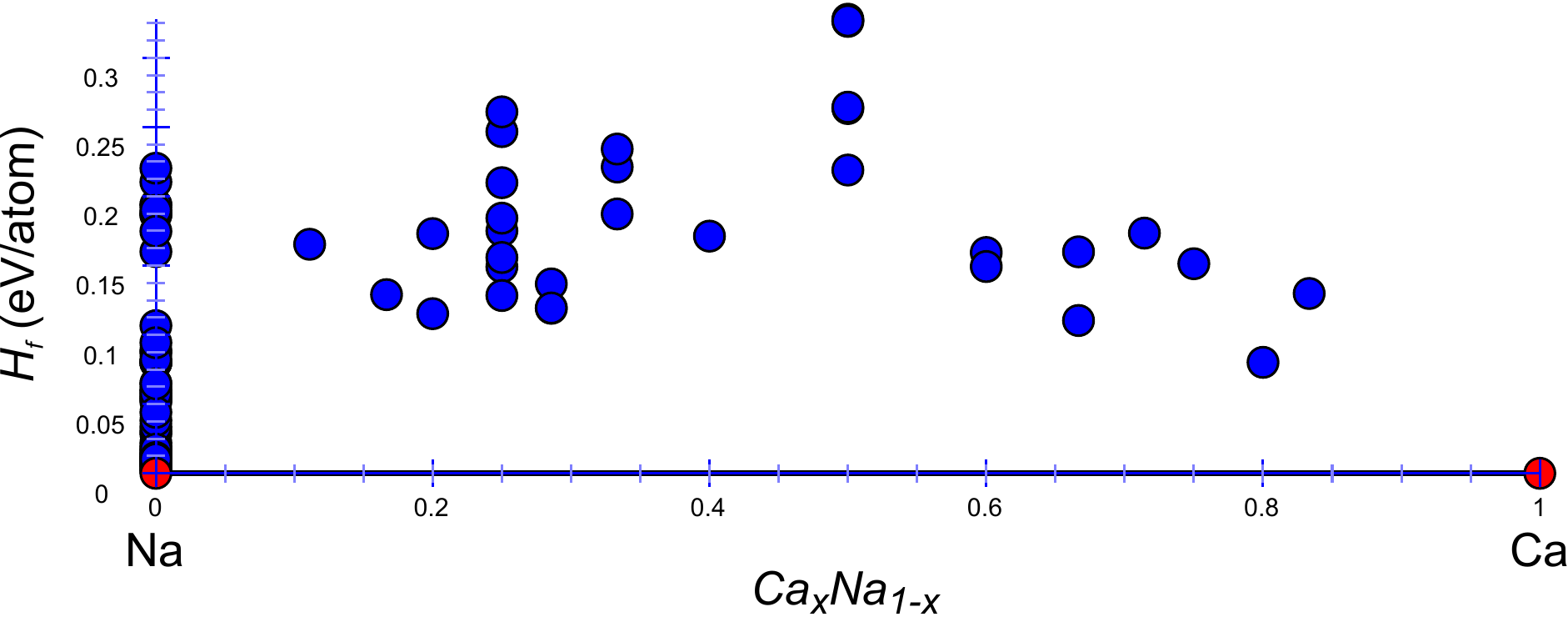}
	\caption[Sodium-Calcium GA Results]{Sodium-calcium binary alloy structure search results. Each point represents  one structure.  Quantity plotted is enthalpy of formation: $\textit{\textsf{h}}_\textit{\textsf{f}}\textsf{(Ca}_\textsf{x}\textsf{Na}_\textsf{1-x}\textsf{)}\equiv \textit{\textsf{h}}\textsf{(Ca}_\textsf{x}\textsf{Na}_\textsf{1-x}\textsf{)}-\textsf{x}\textit{\textsf{h}}\textsf{(Ca)}-\textsf{(1-x)}\textit{\textsf{h}}\textsf{(Na)}$.  Any compound for which $\textit{\textsf{h}}_\textit{\textsf{f}}$ is negative is predicted to be stable.  We found no stable Na-Ca compounds.}
	\vspace{-12pt}
	\label{fig:NaCaGA}
\end{wrapfigure}
The final aspect of our work was an extension of our methods to the sodium-calcium binary alloy system.  We performed a structure search at 50 GPa using crossover and mutation genetic operators, including the BCC phase of sodium and the $\beta$-tin phase of calcium as reference states.  The results of our search after four generations are shown in Fig.~\ref{fig:NaCaGA}.  In order to effectively search the binary phase space, our algorithm requires much more computation time compared to a single element structure search.  Thus, the results presented here are preliminary, but they indicate that our algorithm does fully sample the entire solution space.  One limitation of the binary search is that the structure of each compound may not be fully optimized, and thus the enthalpy may not be fully minimized.  Our search also seemed to converge towards pure sodium in later generations.  Restricting future searches to smaller segments of the composition space and performing structure searches on single compounds would help alleviate these problems.

\section{Conclusions:}

We have performed a computational study of elemental barium and the sodium-calcium binary alloy system.  We used a genetic algorithm to predict stable phases of barium \textit{ab initio} utilizing DFT as implemented within VASP to calculate energies.  We discovered all known phases of barium, excluding incommensurate Ba-IV.  We also discovered a novel structure, $\alpha$-Sm, that has not been previously reported in barium literature, which we predict to be the ground state at moderate pressures.  We improved upon the computational results of Reed and Ackland with respect to Ba-IV, but were not able to predict this phase to be stable without an energy correction. Further DFT investigation of the electronic properties of these structures would prove useful in more fully characterizing the transitions between these phases.  We also recommend an analysis of close packed structures with various stacking orders both in barium and in similar ``simple'' metals.  We were unable to predict any stable compounds in the Na-Ca system, although our search did identify some potential candidate structures with relatively low $\textit{\textsf{h}}_\textit{\textsf{f}}$ values.  Future work will include continued searches at more pressures, as well as structure searches on individual compounds of interest.

Calculations for this work were performed on the Nanolab Computing Cluster at the Cornell Nanoscale Facility.
%,a member of the National Nanotechnology Infrastructure Network supported by the National Science Foundation
The authors would particularly like to thank Derek Stewart for his support with regards to the cluster.  This research used computational resources of the Computational Center for Nanotechnology Innovation at Rensselaer Polytechnic Institute. Finally, this research was supported in part by the National Science Foundation through TeraGrid resources provided by the Texas Advanced Computing Center (TACC) under grant number TG-DMR050028N.

\footnotesize
\bibliographystyle{ieeetr}
\bibliography{Final_manuscript}

\begin{thebibliography}{10}

\bibitem{mourachkine2004}
A.~Mourachkine, {\em {Room-Temperature Superconductivity}}.
\newblock Cambridge International Science Pub., 2004.

\bibitem{leggett2006}
A.~J. Leggett, ``What do we know about high {T$_{\textrm{c}}$}?,'' {\em Nature
  Physics}, vol.~2, no.~3, pp.~134--136, 2006.

\bibitem{degtyareva2010}
O.~Degtyareva, {\em Simple Metals at High Pressures}, pp.~261--280.
\newblock Berlin: Springer Science, 2010.

\bibitem{feng2008}
J.~Feng, R.~G. Hennig, N.~W. Ashcroft, and R.~Hoffmann, ``Emergent reduction of
  electronic state dimensionality in dense ordered {Li-Be} alloys,'' {\em
  Nature}, vol.~451, no.~7177, pp.~445--448, 2008.

\bibitem{xie2010}
Y.~Xie, A.~R. Oganov, and Y.~Ma, ``Novel high pressure structures and
  superconductivity of {CaLi}$_{2}$,'' {\em Phys. Rev. Lett.}, vol.~104,
  p.~177005, Apr 2010.

\bibitem{martin2004}
R.~Martin, {\em Electronic Structure: Basic Theory and Practical Methods}.
\newblock Cambridge University Press, 2004.

\bibitem{nelmes1999}
R.~J. Nelmes, D.~R. Allan, M.~I. McMahon, and S.~A. Belmonte, ``Self-hosting
  incommensurate structure of barium {IV},'' {\em Physical Review Letters},
  vol.~83, no.~20, pp.~4081--4084, 1999.

\bibitem{reed2000}
S.~K. Reed and G.~J. Ackland, ``Theoretical and computational study of
  high-pressure structures in barium,'' {\em Physical Review Letters}, vol.~84,
  no.~24, pp.~5580--5583, 2000.

\bibitem{heine2000}
V.~Heine, ``Crystal structure: As weird as they come,'' {\em Nature}, vol.~403,
  no.~6772, pp.~836--837, 2000.

\bibitem{neaton2001}
J.~B. Neaton and N.~W. Ashcroft, ``On the constitution of sodium at higher
  densities,'' {\em Physical Review Letters}, vol.~86, no.~13, pp.~2830--2833,
  2001.

\bibitem{oganov2010}
A.~R. Oganov, Y.~M. Ma, Y.~Xu, I.~Errea, A.~Bergara, and A.~O. Lyakhov,
  ``Exotic behavior and crystal structures of calcium under pressure,'' {\em
  Proceedings of the National Academy of Sciences}, vol.~107, no.~17,
  pp.~7646--7651, 2010.

\bibitem{CaNaPhase}
A.~Pelton, ``{{Ca} (Calcium) - {Na} (Sodium) Binary Phase Diagram},'' in {\em
  Alloy Phase Diagrams}, \textit{http://www.asmmaterials.info/}: ASM
  International, 2004.

\bibitem{tipton2007}
W.~Tipton, ``A genetic algorithm for crystal structure prediction,'' {\em
  lithium.ccmr.cornell.edu/$\sim$wtipton/gasp/}, 2007.
\newblock \textit{Report available at request from author}.

\bibitem{vasp1}
G.~Kresse and J.~Furthmuller, ``Efficiency of ab-initio total energy
  calculations for metals and semiconductors using a plane-wave basis set,''
  {\em Computational Materials Science}, vol.~6, no.~1, pp.~15--50, 1996.

\bibitem{vasp2}
G.~Kresse and J.~Furthmuller, ``Efficient iterative schemes for ab initio
  total-energy calculations using a plane-wave basis set,'' {\em Physical
  Review B}, vol.~54, no.~16, pp.~11169--11186, 1996.

\bibitem{vasp3}
G.~Kresse and J.~Hafner, ``Ab-initio molecular-dynamics for liquid-metals,''
  {\em Physical Review B}, vol.~47, no.~1, pp.~558--561, 1993.

\bibitem{vasp4}
G.~Kresse and J.~Hafner, ``Ab-initio molecular-dynamics simulation of the
  liquid-metal amorphous semiconductor transition in germanium,'' {\em Physical
  Review B}, vol.~49, no.~20, pp.~14251--14269, 1994.

\bibitem{PBE}
J.~P. Perdew, K.~Burke, and M.~Ernzerhof, ``Generalized gradient approximation
  made simple,'' {\em Phys. Rev. Lett.}, vol.~77, no.~18, p.~3, 1996.

\bibitem{PBEerratum}
J.~P. Perdew, K.~Burke, and M.~Ernzerhof, ``Erratum: Generalized gradient
  approximation made simple,'' {\em Physical Review Letters}, vol.~78, no.~7,
  pp.~1396--1396, 1997.

\bibitem{PBExc}
J.~P. Perdew and A.~Zunger, ``Self-interaction correction to density-functional
  approximations for many-electron systems,'' {\em Physical Review B}, vol.~23,
  no.~10, pp.~5048--5079, 1981.

\bibitem{PP2PAW}
G.~Kresse and D.~Joubert, ``From ultrasoft pseudopotentials to the projector
  augmented-wave method,'' {\em Physical Review B}, vol.~59, no.~3,
  pp.~1758--1775, 1999.

\bibitem{PAW}
P.~E. Blochl, ``Projector augmented-wave method,'' {\em Physical Review B},
  vol.~50, no.~24, pp.~17953--17979, 1994.

\bibitem{ISOTROPY}
H.~Stokes, D.~Hatch, and B.~Campbell, {\em ISOTROPY}.
\newblock \textit{stokes.byu.edu/isotropy.html}, 2007.

\bibitem{navalStructures}
{Center for Computational Materials Science of the United States Naval Research
  Laboratory}, ``Crystal lattice structures.''
  \textit{http://cst-www.nrl.navy.mil/lattice/}, 2011.

\bibitem{arapan2008}
S.~Arapan, H.~K. Mao, and R.~Ahuja, ``Prediction of incommensurate crystal
  structure in ca at high pressure,'' {\em Proceedings of the National Academy
  of Sciences}, vol.~105, no.~52, pp.~pp. 20627--20630, 2008.

\bibitem{chen1988}
Y.~Chen, K.~M. Ho, and B.~N. Harmon, ``First-principles study of the
  pressure-induced bcc-hcp transition in {Ba},'' {\em Phys. Rev. B}, vol.~37,
  pp.~283--288, Jan 1988.

\bibitem{overhauser1984}
A.~W. Overhauser, ``Crystal structure of lithium at 4.2 k,'' {\em Phys. Rev.
  Lett.}, vol.~53, pp.~64--65, Jul 1984.

\bibitem{redfield1986}
A.~C. Redfield and A.~M. Zangwill, ``Stacking sequences in close-packed
  metallic superlattices,'' {\em Phys. Rev. B}, vol.~34, pp.~1378--1380, Jul
  1986.

\end{thebibliography}

\end{document}